\documentclass[pra,twocolumn,showpacs,amsmath,amssymb,floatfix]{revtex4}

\usepackage{graphicx}   
\usepackage{dcolumn}    
\usepackage{bm}         

\def\tthyphen{\discretionary{-}{}{-}}

\input{Qcircuit}

\begin{document}

\title{Optimal classical-communication-assisted local model of
$\bm{n}$-qubit Greenberger-Horne-Zeilinger correlations}

\author{Tracey E. Tessier}

\email{tessiert@info.phys.unm.edu}

\author{Carlton M. Caves}

\author{Ivan H. Deutsch}

\affiliation{Department of Physics and Astronomy, University of New
Mexico, Albuquerque, New Mexico 87131}

\author{Dave Bacon}

\affiliation{Santa Fe Institute, Santa Fe, New Mexico 87501}

\author{Bryan Eastin}

\affiliation{Department of Physics and Astronomy, University of New
Mexico, Albuquerque, New Mexico 87131}

\begin{abstract}
We present a model, motivated by the criterion of reality put forward
by Einstein, Podolsky, and Rosen and supplemented by classical
communication, which correctly reproduces the quantum-mechanical
predictions for measurements of all products of Pauli operators on an
$n$-qubit GHZ state (or ``cat state'').  The $n-2$ bits employed by our
model are shown to be optimal for the allowed set of measurements,
demonstrating that the required communication overhead scales linearly
with $n$.  We formulate a connection between the generation of the
local values utilized by our model and the stabilizer formalism, which
leads us to conjecture that a generalization of this method will shed
light on the content of the Gottesman-Knill theorem.
\end{abstract}

\date{\today{}}


\maketitle

\section{Introduction}

Bell's theorem~\cite{Bell64} codifies the observation that entangled
quantum-mechanical systems exhibit stronger correlations than are
achievable within any local hidden-variable (LHV)  model.  Beyond
philosophical implications, the ability to operate outside the
constraints imposed by local realism serves as a resource for
information processing tasks such as communication~\cite{Schumacher96},
computation~\cite{Nielsen00}, and cryptography~\cite{Ekert91}.

The violation of Bell-type inequalities demonstrates the in-principle
failure of LHV models to account for all of the predictions of quantum
mechanics.  One approach to quantifying the observed difference between
classically correlated systems and entangled states is to translate a
quantum protocol involving entanglement into an equivalent protocol
that utilizes only classical resources, e.g., the shared randomness of
LHV's and ordinary classical communication.  Toner and
Bacon~\cite{Toner03} showed that the quantum correlations arising from
local projective measurements on a maximally entangled state of two
qubits can be simulated exactly using a LHV model augmented by just a
single bit of classical communication.  Pironio~\cite{Pironio03} took
this analysis a step further, showing that the amount of violation of a
Bell inequality imposes a lower bound on the average communication
needed to reproduce the quantum-mechanical correlations.

The original Bell-type inequalities \cite{Bell64,Clauser78} were
formulated for pairs of qubits.  Greenberger, Horne, and
Zeilinger~\cite{Greenberger89} introduced a qualitatively stronger
test of local realism, based on a three-qubit state,
$|\psi_3\rangle=(|000\rangle+|111\rangle)/\sqrt2$, which is now
called the GHZ (or ``cat'') state.  Here $|0\rangle$($|1\rangle$)
represents the eigenvector of the Pauli $Z$ operator with eigenvalue
$+1$($-1$). GHZ correlations have been experimentally demonstrated
in entangled three-photon systems~\cite{Bouwmeester99} and shown to
be useful for performing information-theoretic tasks such as
entanglement broadcasting~\cite{Tong00} and quantum secret
sharing~\cite{Bagherinezhad03}.

Mermin~\cite{Mermin90} introduced a simple argument that shows how
correlations between Pauli operators measured on a GHZ state violate
local realism.  We briefly review Mermin's argument in
Sec.~\ref{Sec:GHZ_Correlations}.  Mermin's formulation is based on
the Einstein-Podolsky-Rosen (EPR)~\cite{Einstein35} reality
criterion: ``If, without in any way disturbing a system, we can
predict with certainty\ \ldots\ the value of a physical quantity,
then there exists an element of physical reality corresponding to
this physical quantity."  This criterion is meant to capture what it
means for a physical system to `possess' a certain property.

Taking the EPR concept of an element of reality as our starting point,
we formulate a LHV model for the $n$-qubit GHZ state.  By itself, the
model is inadequate.  It cannot give the correct quantum-mechanical
predictions for measurements of arbitrary products of Pauli operators
and correlations among such measurements, as is clear from Mermin's
argument and its generalization to $n$ qubits.  Nonetheless, as we show
in Sec.~\ref{Sec:LHV_Model}, when the model is augmented by $n-2$ bits
of classical communication, it does reproduce all the
quantum-mechanical predictions for measurements of Pauli products and
their correlations.  We go on to prove in Sec.~\ref{Sec:Optimality}
that this amount of classical communication is optimal for the allowed
set of measurements, i.e., for measurements of Pauli products.

In Sec.~\ref{Sec:Circuit} we demonstrate that our model arises
naturally from a LHV simulation of a quantum circuit that creates the
$n$-qubit GHZ state.  The quantum circuit consists of an initial
Hadamard gate $H$ followed by a sequence of controlled-NOT (C-NOT)
gates.  It is a special case of a general class of quantum circuits
identified by the Gottesman-Knill (GK) theorem~\cite{Nielsen00}.  The
GK circuits are those composed of qubits (i)~initially prepared in the
state $|00\ldots0\rangle$, (ii)~acted upon by gates in the Clifford
group, which is generated by $H$, $90^\circ$ rotation about $Z$, and
C-NOT~\cite{Gottesman98}, and (iii)~subjected to measurements of
products of Pauli operators.  These circuits are capable of generating
globally entangled states, such as the GHZ state, but their evolution
can nevertheless be simulated in $O(n^{2}/\log n)$ operations on a
classical computer~\cite{Nielsen00,Gottesman98,Aaronson04}.  We return
to the question of GK~simulations in Sec.~\ref{Sec:Conclusion},
comparing and contrasting them with our simulation of the creation of a
GHZ state and speculating on how our results might impact understanding
of the GK theorem.

\section{GHZ Correlations\label{Sec:GHZ_Correlations}}

Mermin's three-qubit GHZ argument can be summarized as follows.  The
three-qubit GHZ state $|\psi_3\rangle$ is uniquely specified as the
simultaneous $+1$ eigenstate of a complete set of commuting Pauli
products, one choice for which is $\langle -XYY,-YXY,-YYX\rangle$,
where the ordering in the product specifies which qubit the Pauli
operator applies to.  In the language of the stabilizer formalism
\cite{Nielsen00,Gottesman98}, the three commuting operators are
referred to as \textit{stabilizer generators\/} of $|\psi_3\rangle$.
The stabilizer generators give the definite outcome $+1$ when measured,
implying that a measurement of two of the Pauli operators in a
generator can be used to predict the result of a measurement of the
third with certainty.  Thus, according to the EPR reality criterion, we
should associate a local element of reality, having value $+1$ or $-1$,
with the $X$ and $Y$ Pauli operators of each qubit.  Letting $x_j$ and
$y_j$ denote the values of these six elements of reality, where $j$
labels the qubit, the stabilizer generators require that
$x_1y_2y_3=y_1x_2y_3=y_1y_2x_3=-1$.  Multiplying these three quantities
together gives $x_1x_2x_3=-1$, showing that the model predicts the
result $-1$ with certainty for a measurement of $XXX$.  Because of the
anticommutativity of the Pauli operators, however, the product of the
stabilizer generators is $+XXX$, showing that quantum mechanics
predicts the result $+1$ for this measurement with certainty.  Mermin's
GHZ argument demonstrates the incompatibility of quantum theory with
local realism.

The $n$-qubit GHZ state,
$|\psi_n\rangle=(|00\ldots0\rangle+|11\ldots1\rangle)/\sqrt2$, is
specified by $n$ stabilizer generators $\langle X^{\otimes n},
ZZI^{\otimes (n-2)}, ZIZI^{\otimes (n-3)},\ldots, ZI^{\otimes (n-2)}
Z\rangle$, where $I$ is the identity operator.  The full
\textit{stabilizer group\/}~\cite{Nielsen00}, generated by these
generators, consists of the $2^n$ commuting Pauli products of which
$|\psi_n\rangle$ is a $+1$ eigenstate; it contains Pauli products that
have (i)~only $I$'s and an even number of $Z$'s and (ii)~only $X$'s and
an even number of $Y$'s, with an overall minus sign if the number of
$Y$'s is not a multiple of 4.  Of the $2\times 4^n$ Pauli products
(including a $\pm$ in front of the product), $2^n$ are members of the
stabilizer group, $2^n$ are negatives of the stabilizer-group elements
and thus yield $-1$ with certainty when measured, and all the rest
return $\pm 1$ with equal probability.

Mermin's argument generalizes straightforwardly to $|\psi_n\rangle$
(our proof of optimality in Sec.~\ref{Sec:Optimality} can be viewed as
just such a generalization) and shows that no local realistic model can
correctly predict the outcomes of all measurements of products of Pauli
operators performed on $|\psi_n\rangle$ and correlations among such
measurements.  We now present a classical-communication-assisted LHV
model that does yield all of the correct quantum-mechanical predictions.

\begin{table}
\begin{tabular}{cccc}
\hspace{50pt}&$\hspace{25pt}X\hspace{25pt}$
&$\hspace{25pt}Y\hspace{25pt}$&$\hspace{25pt}Z\hspace{25pt}$\\
qubit~1&$R_2R_3\cdots R_n$&$iR_1R_2\cdots R_n$&$R_1$\\
qubit~2&$R_2$&$iR_1R_2$&$R_1$\\
qubit~3&$R_3$&$iR_1R_3$&$R_1$\\
$\vdots$&$\vdots$&$\vdots$&$\vdots$\\
qubit~$n$&$R_n$&$iR_1R_n$&$R_1$
\end{tabular}
\caption{Table of LHV's associated with an $n$-qubit GHZ state. Each
row corresponds to a qubit, and each column to a measurement.  The
quantities $R_j$ denote classical random variables that return $\pm 1$
with equal probability. The origin and meaning of the subscripts $j$
becomes clear when we consider the creation of a GHZ state in
Sec.~\protect\ref{Sec:Circuit}. The outcome predicted for a joint
measurement of a Pauli product is obtained by multiplying the
corresponding table entries for each qubit (using 1 for unmeasured
qubits, i.e., for an identity operator appearing in the Pauli product)
and discarding any factor of $i$ in the final product. For example, for
a joint measurement of $XYY$ on the ($n=3$)-qubit GHZ state, our model
predicts the outcome $(R_2R_3)(iR_1R_2)(iR_1R_3)=-1$, in agreement with
quantum mechanics. Here we have used the fact that $R_j^2 = 1$.
Similarly, for a measurement of $IYZ$, the product of table entries is
$(iR_1R_2)(R_1)=iR_2$; with the $i$ discarded, the predicted outcome is
the random result $R_2$, again in accord with quantum mechanics. The
use of $i$ does not mean that the results of $Y$ measurements are
imaginary; rather the $i$ is a ``flag'' that tells us how to combine
$Y$ values for different qubits in a joint measurement. Although the
LHV table might seem not to respect the qubit-exchange symmetry of the
GHZ state, one easily sees that it does by defining $R'_2=R_2\cdots
R_n$, which exchanges the roles of the first and second qubits in the
table. \label{table}}
\end{table}

\section{Communication-Assisted Local Model of GHZ Correlations\label{Sec:LHV_Model}}

Our LHV model is specified in Table~\ref{table}, which lists local
realistic values for the $X$, $Y$, and $Z$ Pauli operators of each
qubit. The caption describes how to determine the predicted outcome for
a measurement of any Pauli product by multiplying the appropriate table
entries and discarding any factor of $i$ in the final product.  The use
of the imaginary phase $i$ in the $Y$ column, apparently just a
curiosity, actually plays a crucial role. It reconciles some of the
conflicting predictions of commuting LHV's and anticommuting Pauli
operators, which form the basis of Mermin's GHZ argument.  More
precisely, the multiplicative algebra of these phases provides a
concise representation of the $n-2$ bits of classical communication
required to ensure that our LHV model yields all of the correct
quantum-mechanical predictions.

To show that Table~\ref{table} gives correct predictions for
measurements of Pauli products, we consider those measurements for
which the table predicts a definite outcome. Suppose first that a Pauli
product contains no $X$'s or $Y$'s, but consists solely of $I$'s and
$Z$'s.  Then it is clear that the table predicts certainty, with the
outcome being $+1$, if and only if the number of $Z$'s in the product
is even.  Suppose now that the product has an $X$ or a $Y$ in the first
position.  Then it is apparent that to avoid a random variable in the
overall product, all of the other elements in the product must be $X$'s
or $Y$'s and the number of $Y$'s must be even; the outcome is $+1$ if
the number of $Y$'s is a multiple of 4 and $-1$ otherwise.  Finally,
suppose the Pauli product has an $X$ or a $Y$ in a position other than
the first.  Then the only way to avoid a random variable in the overall
product is to have an $X$ or a $Y$ in the first position, and we
proceed as before.  These considerations show that our model predicts a
definite outcome, with the correct sign, for precisely those Pauli
products that are in the stabilizer group (or their negatives),
including as a special case the observables forming the basis of
Mermin's GHZ argument.  Likewise, the model correctly predicts a random
result for all other Pauli products.

Our model correctly predicts the outcomes for all measurements of
Pauli products, including single-qubit measurements.  It fails,
however, in some of its predictions for correlations between
single-qubit measurements.  To be correct, the model would have to
reproduce all these correlations for all sets of single-qubit
measurements.  The model fails because products of single-qubit
measurement results predicted by the model are not always equal to
the corresponding joint measurement results.  This inconsistency is
a direct consequence of the rule that discards $i$ from a calculated
measurement outcome.  As an example, consider the single-qubit
measurements $XII$, $IYI$, and $IIY$ on a three-qubit GHZ state. The
product of the single-qubit measurement results, $R_2R_3$ for $XII$,
$R_1R_2$ for $IYI$, and $R_1R_3$ for $IIY$ is $+1$, which is
inconsistent with the prediction of the model and of quantum
mechanics for a joint measurement of the observable $XYY$.  From the
perspective of the model, classical communication between qubits, an
obviously nonlocal element, is necessary precisely to ensure the
consistency of joint measurement predictions with products of
single-qubit predictions.

The inconsistency between joint and correlated local predictions is
a general feature of our LHV model for $n$-qubit GHZ states.  It
occurs only for joint measurements that involve $Y$ measurements on
some qubits and that have a definite outcome, i.e., joint
measurements of stabilizer elements that contain $Y$'s.  Joint
measurements that yield a random result do not suffer from this
problem because the randomness of a product is unaffected by
discarding $i$'s.  More precisely then, the inconsistency occurs
only for joint measurements that are products of $X$'s and $Y$'s on
all the qubits, with the number $Y$'s being an even number that is
not a multiple of 4.

The protocol for ensuring consistency between joint and composite local
predictions proceeds as follows.  An observer called Alice, stationed
at, say, the first qubit, is put in charge of ensuring consistency with
single-qubit measurements.  She does so by changing or not changing the
sign of the outcome on her qubit, based on what is measured on her
qubit and information she receives about what is measured on the other
qubits.  Because of the qubit-exchange symmetry of the GHZ state (and
of the LHV table), an observer stationed at any qubit could play the
role of Alice.  Alice ensures consistency by changing the sign of her
local outcome if and only if (i)~a measurement of $X$ or $Y$ is made on
her qubit and (ii)~the total number of $Y$ measurements on all qubits
is an even number that is not a multiple of 4.  The protocol requires
$n-1$ bits of communication as each of the other qubits reports to
Alice whether $Y$ was measured on that qubit.  The protocol clearly
fixes all those cases that need correction; just as important, in all
situations where Alice flips her qubit, all subsets of qubits that
include Alice's qubit, except for the case of a needed correction, have
a random measurement product, which is therefore unaffected by Alice's
flip.  The success of this protocol clearly relies on very special
properties of the stabilizer group for the $n$-qubit GHZ state.

We can put the protocol in a more mathematical form by letting $r_1=1$
if an $X$ or $Y$ measurement is made on the first qubit and $r_1=0$
otherwise and by letting $q_j=i$ if $Y$ is measured on the $j$th qubit
and $q_j=1$ otherwise. Alice ensures consistency by flipping her local
outcome if and only if $p_n=r_1q_1\cdots q_n=-1$.  This formulation
allows us to see easily that we can do a bit better than the $n-1$ bits
required by the original protocol.  The key is to notice that when
$p_n=\pm i$, all subsets of qubits that include Alice's qubit have a
random measurement product, so a flip by Alice goes unnoticed.  As a
result, Alice can get by with the truncated product
$p_{n-1}=r_1q_1\cdots q_{n-1}$, flipping her local outcome if and only
if $p_{n-1}=i$ or $-1$.  This scheme requires the promised $n-2$ bits
of communication, because Alice doesn't need to know whether a $Y$
measurement is made on the $n$th qubit; it works because Alice flips
whenever $p_n=-1$, as required, with the additional flips when $p_n=\pm
i$ not doing any harm.

The consistency scheme generalizes trivially to the case of
Pauli-product measurements made on $l$ disjoint sets of qubits.  For
each set $k$ chosen from the $l$ sets, the table yields a measurement
product that is the predicted outcome multiplied by $q_k=i$ or $q_k=1$.
Putting Alice in charge of the first set, all but the last of the other
sets communicates $q_k$ to Alice, who computes the product
$r_1q_1\cdots q_{l-1}$, where $r_1=0$ if no measurement or a $Z$
measurement is made on any qubit in her set and $r_1=1$ otherwise.
Alice flips her set's outcome if and only if $r_1q_1\cdots q_{l-1}=i$
or $-1$.  Consistency is thus ensured at the price of $l-2$ bits of
communication.

\section{Proof of Optimality\label{Sec:Optimality}}

Using an elaboration of Mermin's GHZ argument~\cite{Mermin90}, we now
demonstrate that our model is optimal by showing that any protocol that
is allowed at most $n-3$ bits of classical communication is incapable
of yielding all quantum-mechanical predictions for measurements of
Pauli products on $|\psi_n\rangle$ and their correlations.  For this
purpose, imagine the $n$ qubits as the nodes of a graph; two qubits are
connected by a line if at least one bit is communicated between them.
The graph partitions the qubits into disjoint connected subsets.  There
being at most $n-3$ lines, it follows that there are at least three
disconnected subsets, since at best each line consolidates two subsets
into one, thereby eliminating one subset.  Moreover, it is always
possible to arrange the communication so that there are three subsets.
The communication can do no more than allow us to treat all Pauli
products within a subset as a single joint observable.  We can restrict
attention to the case of three subsets, since amalgamating disconnected
subsets allows the communication more power than it actually has.

The situation then is that we have three disconnected subsets,
containing $k$, $l$, and $m$ qubits, with $k+l+m=n$.  Since the GHZ
state is invariant under qubit exchange, we can make the first $k$
qubits those in the first subset and the next $l$ qubits those in the
second subset, leaving the final $m$ qubits to be those in the third
subset.  We now define six Pauli products: $A=X^{\otimes(k-1)}X$ and
$B=X^{\otimes(k-1)}Y$ for the first subset, $C=X^{\otimes(l-1)}X$ and
$D=X^{\otimes(l-1)}Y$ for the second subset, and $E=X^{\otimes(m-1)}X$
and $F=X^{\otimes(m-1)}Y$ for the third subset.  The four operators,
$ACE$, $-ADF$, $-BCF$, $-BDE$, are in the stabilizer group of
$|\psi_n\rangle$ and thus give a definite outcome $+1$ when measured,
implying that a measurement of any two of the operators in the product
can be used to predict with certainty the result of a measure of the
third.  The EPR reality criterion then says that we should associate
elements of reality, having values $\pm1$, with the six Pauli products
$A$-$F$.  Denoting the values of these elements of reality in the
obvious way, the definite values of the last three stabilizer elements
imply that $adf=bcf=bde=-1$.  The product of these three quantities is
$ace=-1$, contradicting the $+1$ prediction of quantum mechanics for a
measurement of $ACE$.

For completeness, we note another form of the argument.  According
to the elements of reality, the observable ${\cal
M}=ACE-ADF-BCF-BDE$ has the value
\begin{equation}
{\cal M}=ace-adf-bcf-bde=c(ae-bf)-d(af+be)\;.
\end{equation}
Since $ae=\pm bf\Longleftrightarrow af=\pm be$, it is easy to see that
${\cal M}=\pm2$.  This implies that the expectation value satisfies
$|\langle{\cal M}\rangle|\le2$, whereas the $n$-qubit GHZ state has
$\langle{\cal M}\rangle=4$.  This form of the argument does not make
use of the properties of the GHZ state, and it makes clear that
stochastic models can do no better than the deterministic models
considered here.  We also note that this argument produces a Bell
inequality with auxiliary communication\cite{Bacon03}.

\section{Quantum Circuit\label{Sec:Circuit}}

\begin{figure}
\hspace{-100pt} \Qcircuit @C=1.4em @R=1.6em {
    \lstick{\ket{0}} & \gate{H} & \ctrl{1}  & \ctrl{2}
    & \qw \\
    \lstick{\ket{0}} & \qw      & \targ     & \qw
    &  \qw & \rstick{\displaystyle{\ket{\psi_{3}}=
        \frac{1}{\sqrt{2}}\left(\ket{000}+\ket{111}\right)}}\\
    \lstick{\ket{0}} & \qw      & \qw       & \targ
    & \qw \\
} \caption{Quantum circuit that generates the three-qubit GHZ
state.} \label{fig:GHZ}
\end{figure}
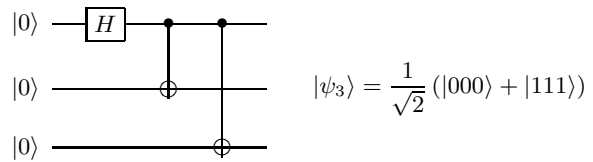

Table~\ref{table} is the basis of our communication-assisted LHV model.
It arises naturally from a quantum circuit that creates the $n$-qubit
GHZ state from an initial state $|00\ldots 0\rangle$.  One such circuit
consists of a Hadamard gate on the first qubit followed by $n-1$ C-NOT
gates, with the leading qubit being the control and the remaining
qubits serving successively as targets. The three-qubit version of this
circuit is shown in Fig.~\ref{fig:GHZ}. The Hadamard gate $H$
transforms the Pauli operators according to
\begin{equation}
HXH^\dagger=Z,\quad HYH^{\dagger}=-Y,\quad HZH^{\dagger}=X\;.
\label{eq:Htrans}
\end{equation}
Similarly, under the action of C-NOT, we have
\begin{eqnarray}
C(XI)C^\dagger=XX\;,
&C(YI)C^\dagger=YX\;,
&C(ZI)C^\dagger=ZI\;,\nonumber\\
C(IX)C^\dagger=IX\;,
&C(IY)C^\dagger=ZY\;,
&C(IZ)C^\dagger=ZZ\;,\nonumber\\
\label{eq:Ctrans}
\end{eqnarray}
where the first qubit is the control and the second is the target.
These operator transformations lead to the table update rules given
in Fig.~\ref{fig:LHV}, which traces the evolution of the LHV table
during the creation of a three-qubit GHZ state using the circuit of
Fig.~\ref{fig:GHZ}.  A simple generalization to $n$ qubits leads to
Table~\ref{table} for the $n$-qubit GHZ state.

The C-NOT update rules given in Fig.~\ref{fig:LHV} must be consistent
with the fifteen nontrivial transformations of Pauli products generated
by $C$.  Six of these transformations, listed in Eq.~(\ref{eq:Ctrans}),
serve as the basis for the update rules.  Because $C=C^\dagger$, the
rules are automatically consistent with four other transformations. In
addition, the rules are clearly consistent with the transformation
$C(ZX)C^\dagger=ZX$.  Consistency with the remaining four
transformations, $C \left( X Y \right) C^{\dagger} = Y Z$, $C \left( X
Z \right) C^{\dagger} = - Y Y$, and their inverses, requires that
\begin{eqnarray}
X_c^{\rm I}Y_t^{\rm I}&=&Y_c^{\rm F}Z_t^{\rm F}
=Y_c^{\rm I}Z_c^{\rm I}Z_t^{\rm I}X_t^{\rm I}\;,\nonumber\\
X_c^{\rm I}Z_t^{\rm I}&=&-Y_c^{\rm F}Y_t^{\rm F}
=-Y_c^{\rm I}Z_c^{\rm I}X_t^{\rm I}Y_t^{\rm I}\;.
\label{eq:Cdifficulty}
\end{eqnarray}
These relations do not hold generally, but are satisfied if the initial
entries for both the control and target are correlated according to
$XYZ=i$ (or $XYZ=-i$), with $X$ and $Z$ real and $Y$ imaginary.  These
conditions hold in all our applications of C-NOT.  It is for this
reason that the initial sign of the $Y$ entry for the first qubit (see
the first table in Fig.~\ref{fig:LHV}) is opposite that of the
remaining qubits.

\begin{figure}
$
\begin{array}{cccc}
X & Y & Z \\
R_{1} & -iR_{1} & 1 \\
R_{2} & iR_{2} & 1 \\
R_{3} & iR_{3} & 1
\end{array}
\xrightarrow{H_{1}}
\begin{array}{cccc}
X & Y & Z\\
1 & iR_{1} & R_{1}\\
R_{2} & iR_{2} & 1\\
R_{3} & iR_{3} & 1
\end{array}
\xrightarrow{\mbox{\scriptsize{C-NOT}$_{12}$}}
\begin{array}{cccc}
X & Y & Z\\
R_{2} & iR_{1}R_2 & R_{1}\\
R_{2} & iR_{1}R_2 & R_{1}\\
R_{3} & iR_{3} & 1
\end{array}
\vspace{6pt}
\newline
\xrightarrow{\mbox{\scriptsize{C-NOT}$_{13}$}}
\begin{array}{cccc}
X & Y & Z\\
R_{2}R_3 & iR_{1}R_2R_3 & R_{1}\\
R_{2} & iR_{1}R_2 & R_{1}\\
R_{3} & iR_{1}R_3 & R_{1}
\end{array}
$
\caption{Evolution of the LHV table during the creation of a
three-qubit GHZ state using the circuit of Fig.~\protect\ref{fig:GHZ}.
The initial table yields the correct quantum predictions for the state
$|000\rangle$.  The rules for updating the table through Hadamard and
C-NOT gates come from the operator
transformations~(\protect\ref{eq:Htrans}) and
(\protect\ref{eq:Ctrans}).  The Hadamard update rules are
\leftline{}
\centerline{
$X^{\rm F}=Z^{\rm I}\;,\quad
Y^{\rm F}=-Y^{\rm I}\;,\quad
Z^{\rm F}=X^{\rm I}\;,$
}
\leftline{}
where I and F denote the initial and final values of a table entry,
before and after the application of the gate.  The rules for updating
through a C-NOT, with control~$c$ and target~$t$, are
\leftline{}
\centerline{$
X_c^{\rm F}=X_c^{\rm I}X_t^{\rm I}\;,\quad
Y_c^{\rm F}=Y_c^{\rm I}X_t^{\rm I}\;,\quad
Z_c^{\rm F}=Z_c^{\rm I}\;, $}
\centerline{$
X_t^{\rm F}=X_t^{\rm I}\;,\quad
Y_t^{\rm F}=Z_c^{\rm I}Y_t^{\rm I}\;,\quad
Z_t^{\rm F}=Z_c^{\rm I}Z_t^{\rm I}\;.$}
\leftline{}
The update rules are local in that they only require changes to table
entries corresponding to the qubits involved in a gate.  The subscripts
on the random variables in Table~\protect\ref{table} are now seen to
represent the qubits to which these variables were initially
associated.  Correlations arising from the application of C-NOT gates
then correspond to pairs of identical subscripts.}
\label{fig:LHV}
\end{figure}

\section{Conclusion\label{Sec:Conclusion}}

We have shown that it is possible to reproduce correctly the
quantum-mechanical measurement predictions for the set of all $n$-fold
products of Pauli operators on an $n$-qubit GHZ state using only
Mermin-type LHV's and $n-2$ bits of classical communication. The $n-2$
bits of communication, shown here to be optimal, are required to ensure
that the products of local measurement predictions are consistent with
the corresponding joint predictions. We also show how our model arises
naturally from a simulation of a quantum circuit that creates the
$n$-qubit GHZ state.

The circuit that creates the GHZ state is an example of the
Gottesman-Knill circuits mentioned in the Introduction.   Although GK
circuits can produce global entanglement, as in the $n$-qubit GHZ
state, they can be efficiently simulated in $O(n^{2}/\log n)$ steps on
a classical computer \cite{Nielsen00,Gottesman98,Aaronson04}.  The
important difference between the GK simulation of the circuit that
creates the $n$-qubit GHZ state and our communication-assisted LHV
simulation is that the GK algorithm tracks the evolution of
\textit{nonlocal\/} hidden variables, represented by the $n$ generators
of the stabilizer group, and therefore requires no communication
overhead to give correct predictions.

A generalization of our results might lead to a new perspective on the
GK theorem since our results imply that, at least in this limited case,
we can replace the nonlocal hidden variables represented by the
stabilizer generators with LHV's and a linear amount of classical
communication.  We conjecture that this is a generic feature of quantum
circuits obeying the constraints of the GK theorem; that is, we expect
that any quantum state produced by a GK circuit can be modelled with
EPR elements of reality plus an amount of classical communication that
scales linearly in the number of qubits.

There are two main obstacles to a straightforward extension of our LHV
model to general GK circuits.  The first is the difficulty, expressed
in Eq.~(\ref{eq:Cdifficulty}), in maintaining the consistency
conditions for the C-NOT update rules.  The second is the reliance of
our communication protocol on very special properties of the GHZ state;
for general GK states, the communication protocol will have to be more
complicated, with a proof of optimality correspondingly more
complicated as well.  Nonetheless, the existence of a
communication-assisted LHV model for arbitrary GK circuits and the
entangled states they produce is currently under investigation.

\begin{acknowledgments}
We thank R.~Raussendorf for helpful discussions.  The quantum circuit
in Fig.~\ref{fig:GHZ} was set using the \LaTeX\ package {\tt Qcircuit},
available at http://info.phys.unm.edu/Qcircuit/. This work was partly
supported by ARO Grant No.~DAAD19-01-1-0648.
\end{acknowledgments}

\end{document}